# Cooperative spontaneous emission of three identical atoms

Wei Feng, Yong Li, and Shi-Yao Zhu

*Beijing Computational Science Research Center, Beijing 100084, China*

We study the influence of the distribution of atoms on the cooperative spontaneous emission by a simple model of three identical atoms. The effects of counter rotating terms are included by a unitary transformation method. By discussing two special cases that the three atoms are arranged as an equilateral triangle and in a straight line, we find that the superradiance of the coherent system largely dependent on the homogeneity of the atoms' distribution. If the atoms distribute symmetrically, the superradiant emission will be enhanced. Next, we calculate the emission spectra of three identical atoms under the single-photon state. We find that the distribution of atoms also has a great impact on the lamb shift and the spectrum. If three atoms are placed into an equilateral triangle and the same dipole moment is perpendicular to the plane of the three atoms, the spectrum will degenerate into two peaks from the three peaks for general case.

## Ⅰ. INTRODUCTION

The spontaneous emission spectra and the lamb shift of independent atom once stimulated the development of the quantum theory and the quantum electrodynamics. For the spontaneous emission from coherent systems, it was first investigated by Dicke in 1954 [1], and he found that when the dimension of atom's cloud is much smaller than the involved emission wavelength, the coherence between atoms is so strong that it can greatly enhance or suppress the collective spontaneous emission, called the superradiance and subradiance, respectively. If the atom cloud is not very smaller than the wavelength, the distance between the atoms will play an important role in the collective effect. The simplest model is the system of two two-level atoms, which was first investigated in 1969 [2] under the rotating wave approximation (RWA). Later, the collective spontaneous emission was studied with exact integration of the Heisenberg equation [3] or with the master equation method where the counter rotating terms were taken into account [4]. Recently, by using a unitary transformation method [5], the decay and Lamb shift of two multi-level atoms was studied [6].

Recently, the theoretical study of cooperative spontaneous emission of $N$ atoms has attracted renewed interest [7-12], where the electromagnetic field is treated as a scalar field to simplify the calculation. Meanwhile, the research of collective lamb shift has made great progress in experiment [13, 14]. It is well know that the counter rotating terms are must be considered in obtaining the correct lamb shift, and the research in Ref. [10] found that the virtual processes from the counter rotating terms could also have large influence in the time evolution of coupled atoms ensemble. The study of cooperative spontaneous emission of $N$ atoms with including the counter rotating terms can be treated as a many-body eigenvalue problem [15], where the EM field is treated as vector field, the counter rotating terms are included by using the unitary transformation method[5]. In Ref. [15], the $N$ atoms are randomly distributed in cubes, spheres or quasi-two-dimensional square sheets, and the dependence of the emission dynamics on the geometric shape of the ensemble was studied.

The condition of Dicke superradiance, that is, the dimension of the atomic size is much smaller than the typical wavelength: $R \ll \lambda$, is hard to meet. Due to the recent development on the similarity between semiconductor quantum dots and two-level atoms [16], the study of the cooperative effects can be carried on in the semiconductor quantum dots and the superradiance of this system has been observed [17]. The quantum dots as a kind of artificial atoms, their distribution can be easily controlled. There is still lack of the detailed analysis on the influence of the atoms' distribution on the cooperative spontaneous emission and Lamb shift.

In this paper, we study the influence of atom's distribution on the cooperative spontaneous emission and Lamb shift with the model of three identical atoms. In Sec. II, the model and Hamiltonian will be given. We consider in Sec. III the dynamical evolution of the total system of atoms and vacuum fields and obtain the effective "exponentially-decaying" eigenmodes of the atomic system for general case. In Sec. IV, the detailed calculation about the cooperative decay rate and Lamb shift of the atomic system is considered for two special cases: in the first case the three atoms are arranged as an equilateral triangle with all the dipope moments normal to the plane of triangle, and in the second case the three atoms are arranged in a straight line. In Sec. V, we calculate the spontaneous emission spectra of an initial state with only one atom being excited for the three-atom system in both cases of equilateral triangle and straight line in order to consider the influence of atomic distribution on the spectrum. At last, a brief conclusion will be given in Sec. VI.

## II. MODEL AND HAMILTONIAN

We consider the system consisted of three identical multi-level atoms interacting with the vacuum field. The atoms are fixed in space, and labeled by 1, 2 and 3, respectively. The Hamiltonian of the system can be written as ($\hbar=1$) [18]:

$$H = H_0 + H_I + H_{es},\tag{1}$$

$$H_0 = \sum_n \sum_i \omega_i |i\rangle_n \langle i|_n + \sum_{\mathbf{k}} \omega_k b_{\mathbf{k}}^\dagger b_{\mathbf{k}},\tag{2}$$

$$H_I = \sum_{n,\mathbf{k}} \sum_{j \neq i} g_{\mathbf{k},ij} |i\rangle_n \langle j|_n (b_{\mathbf{k}}^\dagger e^{-i\mathbf{k}\cdot\mathbf{r_n}} + b_{\mathbf{k}} e^{i\mathbf{k}\cdot\mathbf{r_n}}),\tag{3}$$

$$H_{es} = \frac{1}{4\pi\varepsilon_0} \sum_{m<n} \left[ \frac{\mathbf{d}^{(m)} \cdot \mathbf{d}^{(n)}}{r_{mn}^3} - \frac{3(\mathbf{d}^{(m)} \cdot \mathbf{r}_{mn})(\mathbf{d}^{(n)} \cdot \mathbf{r}_{mn})}{r_{mn}^5} \right],\quad (m, n=1,2,3).\tag{4}$$

$H_0$ is the unperturbed Hamiltonian of the atoms and vacuum field, $H_I$ is the interaction Hamiltonian between the atoms and the vacuum modes, and $H_{es}$ is the electrostatic dipole-dipole interaction Hamiltonian. In above, $\omega_i$ is the energy of level $|i\rangle$, $b_{\mathbf{k}}^\dagger$ ($b_{\mathbf{k}}$) is the creation (annihilation) operator of the **k**th mode with frequency $\omega_k$, and

$g_{\mathbf{k},ij} = \omega_{ij} d_{ij} (2\varepsilon_0 \omega_k V)^{-1/2} \hat{e}_{\mathbf{k}} \hat{\mathbf{d}}_{ij}$ is the coupling strength between the $\mathbf{k}$th EM mode with unit polarization vector $\hat{e}_{\mathbf{k}}$ and the atomic transition between levels $|i\rangle$ and $|j\rangle$ with transition dipole moment $\mathbf{d}_{ij} = e\langle i|\mathbf{r}|j\rangle = d_{ij}\hat{\mathbf{d}}_{ij}$, of which $d_{ij}$ (assumed to be real) and $\hat{\mathbf{d}}_{ij}$ are the magnitude and unit vector, respectively. The $n$th atom is located at position $\mathbf{r}_n$ and its displacement is $\mathbf{r}_{mn} \equiv \mathbf{r}_n - \mathbf{r}_m \equiv r_{mn}\hat{\mathbf{r}}_{mn}$. In Eq. (4), $\mathbf{d}^{(n)} = \sum_{ij} \mathbf{d}_{ij} |i\rangle_n \langle j|_n$ is the dipole moment of the $n$th atom. We assume that the three atoms have the same dipole direction. The angle between the dipole and the vector $\mathbf{r}_{mn}$ is $\eta_{mn}$, as shown in Fig. 1.

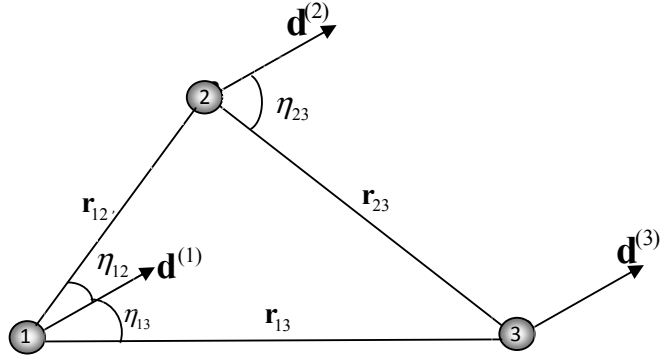

Fig. 1. The schematic description of the relation between dipoles and displacements

In order to take into account the counter rotating terms and simplify the calculation, we introduce a unitary transformation $U = \exp(iS)$ with:

$$S = \sum_{n,\mathbf{k}} \sum_{j \neq i} \frac{g_{\mathbf{k},ij} \xi_{k,ij}}{i\omega_k} |i\rangle_n \langle j|_n (b_{\mathbf{k}}^{\dagger} e^{-i\mathbf{k}\cdot\mathbf{r}_n} - b_{\mathbf{k}} e^{i\mathbf{k}\cdot\mathbf{r}_n}), \qquad (5)$$

where $\xi_{k,ij} = \dfrac{\omega_k}{\omega_k + |\omega_{ji}|}$ and $\omega_{ji} = \omega_j - \omega_i$. By subtracting the free electron self-energy $E_{se} = -\sum_{n,\mathbf{k}} \sum_{j \neq i} \dfrac{|g_{\mathbf{k},ij}|^2}{\omega_k} |i\rangle_n \langle i|_n$, the effective Hamiltonian after the transformation can be written as

$$H^S = e^{iS} H e^{-iS} - E_{se} = H_0 + (H_0' - E_{se}) + H_I' + H_{iv} + H_{es} + o(g_{\mathbf{k},ij}^2), \qquad (6)$$

where

$$H_0' - E_{se} = \sum_{n,\mathbf{k}} \sum_{j \neq i} \frac{|g_{\mathbf{k},ij}|^2}{\omega_k} (\xi_{k,ij}^2 + \frac{\omega_{ji}}{\omega_k}\xi_{k,ij}^2 - 2\xi_{k,ij} + 1)|i\rangle_n \langle i|_n, \tag{7}$$

$$H_I' = \sum_{n,\mathbf{k}} \sum_{i<j} V_{\mathbf{k},ij}(|i\rangle_n \langle j|_n b_\mathbf{k}^\dagger e^{-i\mathbf{k}\cdot\mathbf{r_n}} + |j\rangle_n \langle i|_n b_\mathbf{k} e^{i\mathbf{k}\cdot\mathbf{r_n}}), \tag{8}$$

$$H_{iv} = -\sum_{m \neq n} \sum_\mathbf{k} \frac{2g_{\mathbf{k},eg}^2 \xi_{k,eg}}{\omega_k}(2-\xi_{k,eg}) e^{i\mathbf{k}\cdot\mathbf{r}_{mn}} |e_m\rangle\langle e_n|. \tag{9}$$

Here $V_{\mathbf{k},ij} = g_{\mathbf{k},ij} \frac{2|\omega_{ji}|}{\omega_k + |\omega_{ji}|}$ is the transformed coupling strength, and $|e_n\rangle$ stands for the state that the $n$th atom in the excited state and the others are in their ground state. In the above equations, $H_0' - E_{se}$ only contains the diagonal energy corrections, which is called the non-dynamic shift for a single atom, $H_I'$ is the transformed interaction Hamiltonian between the atoms and the vacuum modes, and $H_{iv}$ is the interaction Hamiltonian due to exchange of virtual photons.

### III. DYNAMIC EVOLUTION

In the transformed total Hamiltonian $H^S$, $H_0 + (H_0' - E_{se}) := H_0^S$ can be rewrite as

$$H_0^S = \sum_n \sum_i \omega'_i |i\rangle_n \langle i|_n + \sum_\mathbf{k} \omega_k b_\mathbf{k}^\dagger b_\mathbf{k}, \tag{10}$$

where $\omega'_i = \omega_i + \delta_i^{nd}$ with $\delta_i^{nd}$ the non-dynamic shift for a single atom

$$\delta_i^{nd} = \sum_{j \neq i,\mathbf{k}} \frac{|g_{\mathbf{k},ij}|^2}{\omega_k} (\xi_{k,ij}^2 + \frac{\omega_{ji}}{\omega_k}\xi_{k,ij}^2 - 2\xi_{k,ij} + 1). \tag{11}$$

In the interaction picture with respect to $H_0^S$, the interaction Hamiltonian becomes

$$H_{IP} = \exp[iH_0^S t] \cdot (H_I' + H_{iv} + H_{es}) \cdot \exp[-iH_0^S t] = H_I^{IP} + H_{sta}^{IP}, \tag{12}$$

where

$$H_I^{IP} = \sum_{n,\mathbf{k}} V_{\mathbf{k},eg} (|g\rangle_n \langle e|_n e^{-i(\omega'_{eg}-\omega_k)t} b_\mathbf{k}^\dagger e^{-i\mathbf{k}\cdot\mathbf{r_n}} + |e\rangle_n \langle g|_n e^{i(\omega'_{eg}-\omega_k)t} b_\mathbf{k} e^{i\mathbf{k}\cdot\mathbf{r_n}}), \quad (13)$$

$$H_{sta}^{IP} = H_{iv}^{IP} + H_{es}^{IP} = H_{iv} + H_{es} = \sum_{m<n} \Delta_{sta}^{mn} (|e_m\rangle\langle e_n| + |e_n\rangle\langle e_m|). \quad (14)$$

Here the shift $\Delta_{sta}^{mn}$ contains the electrostatic dipole-dipole interaction and the interaction due to exchange of virtual photon. It can be written as

$$\Delta_{sta}^{mn} = \frac{1}{4\pi\varepsilon_0} \left[ \frac{\mathbf{d}_{eg}^{(m)} \cdot \mathbf{d}_{ge}^{(n)}}{r_{mn}^3} - \frac{3(\mathbf{d}_{eg}^{(m)} \cdot \mathbf{r}_{mn})(\mathbf{d}_{ge}^{(n)} \cdot \mathbf{r}_{mn})}{r_{mn}^5} \right] - \sum_\mathbf{k} \frac{2 g_{\mathbf{k},eg}^2 \xi_{k,eg}}{\omega_k} (2-\xi_{k,eg}) e^{i\mathbf{k}\cdot\mathbf{r_{mn}}}. \quad (15)$$

We consider the special case that the system in the subspace of single-atom-excitation states. In the interaction picture, the wave function at time *t* can be written as:

$$|\psi(t)\rangle = \sum_n C_n(t)|e_n;0\rangle + \sum_\mathbf{k} C_\mathbf{k}(t)|G;1_\mathbf{k}\rangle, \quad (16)$$

where the state $|G;1_\mathbf{k}\rangle$ represents the three atoms in the ground state and a single photon in mode **k**, and $|e_n;0\rangle$ stands for the state that only the *n*th atom is in its excited level and the electromagnetic field is in the vacuum state.

From the Schrodinger equation $i\partial_t |\psi(t)\rangle = H_{IP} \cdot |\psi(t)\rangle$, we obtain the equations of motion for the state amplitudes in Eq. (16),

$$\dot{C}_n(t) = -i\sum_\mathbf{k} e^{i(\omega'_{eg}-\omega_k)t} V_{\mathbf{k},eg} e^{i\mathbf{k}\cdot\mathbf{r_n}} \cdot C_\mathbf{k}(t) - i\sum_{m\neq n} \Delta_{sta}^{mn} \cdot C_m(t), \quad (17)$$

$$\dot{C}_\mathbf{k}(t) = -i\sum_n e^{-i(\omega'_{eg}-\omega_k)t} V_{\mathbf{k},eg} e^{-i\mathbf{k}\cdot\mathbf{r_n}} \cdot C_n(t). \quad (18)$$

Formally integrating Eq. (18) with the initial value $C_\mathbf{k}(0)=0$, and extending the lower bound of time integration to $-\infty$ under the Weisskopf-Wigner approximation at the long time limit, and then substituting it into Eq. (17), with replacing $C_n(t')$ in the time integration by $C_n(t)$ under the Markov approximation [15], we find (see the Appendix for the details)

$$\dot{C}_n(t) = -\Gamma_0 \cdot C_n(t) - \sum_{m \neq n} \Gamma_{mn} \cdot C_m(t), \tag{19}$$

where

$$\Gamma_0 = \frac{\gamma_{eg}}{2} + i\Delta_{eg}, \tag{20}$$

$$\Gamma_{mn} = \frac{\gamma_{eg}}{2}[D(x_{mn},\eta_{mn}) + i \cdot P(x_{mn},\eta_{mn})], \tag{21}$$

where $\gamma_{eg} = \frac{d_{eg}^2 \omega_{eg}^3}{3\pi\varepsilon_0 c^3}$ is the single-atom decay rate from $|e\rangle$ to $|g\rangle$, $\Delta_{eg}$ is the dynamic shift of the excited state of a single atom, $x_{mn} = \frac{r_{mn}}{\lambda_0}$ ($\lambda_0$ is the wavelength of the resonantly emitted photon), and $\eta_{mn}$ is the angle between the dipole moment and the position vector $\mathbf{r}_{mn}$. The two dimensionless functions, $D(x,\eta)$ and $P(x,\eta)$, are

$$D(x,\eta) = \frac{3}{2}\left\{\sin^2\eta \cdot \frac{\sin(2\pi x)}{2\pi x} + (1 - 3\cos^2\eta)\left[\frac{\cos(2\pi x)}{(2\pi x)^2} - \frac{\sin(2\pi x)}{(2\pi x)^3}\right]\right\}, \tag{22}$$

$$P(x,\eta) = \frac{3}{2}\left\{-\sin^2\eta \cdot \frac{\cos(2\pi x)}{2\pi x} + (1 - 3\cos^2\eta)\left[\frac{\sin(2\pi x)}{(2\pi x)^2} + \frac{\cos(2\pi x)}{(2\pi x)^3}\right]\right\}. \tag{23}$$

In Fig. 2, we plot the two dimensionless functions versus $x$ for different angle $\eta$, where we can see that both of them depend on the dipole direction. Especially, $P(x,\eta)$ is divergent as $x^{-3}$ when $x \to 0$, and it change from $-\infty$ to $+\infty$ when $\eta$ increases, which can be explained by the model of classic dipole antenna [5]. Note that $D(x,\eta)$ and $P(x,\eta)$ oscillate with largest amplitude at the angle of $\eta = \frac{\pi}{2}$, as the radiation is strongest in the direction perpendicular to the dipole.

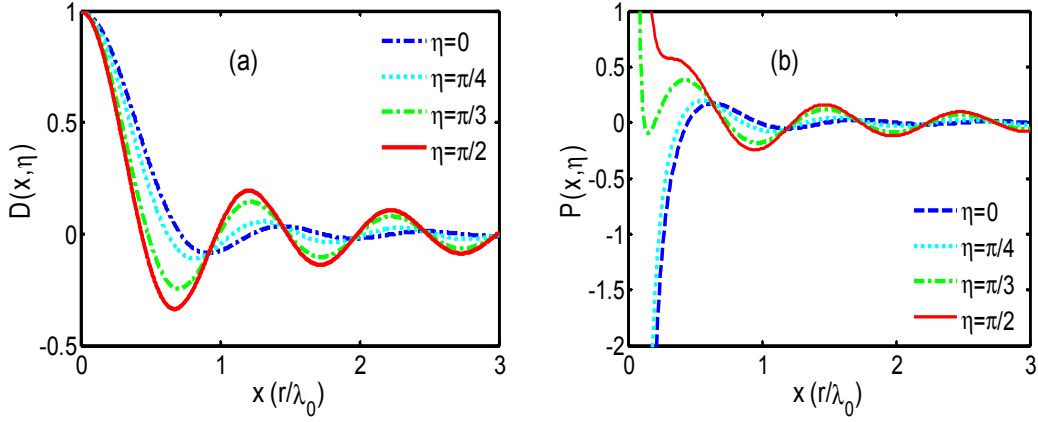

Fig. 2. (Color online) (a) $D(x,\eta)$ and (b) $P(x,\eta)$ versus $x$ with different $\eta$.

The diagonal element $\Gamma_0$ and the non-diagonal elements $\Gamma_{mn}$ constitute a symmetric $3\times 3$ matrix, $\mathbf{\Gamma}$,

$$\mathbf{\Gamma} = \begin{bmatrix} \Gamma_0 & \Gamma_{12} & \Gamma_{13} \\ \Gamma_{12} & \Gamma_0 & \Gamma_{23} \\ \Gamma_{13} & \Gamma_{23} & \Gamma_0 \end{bmatrix}, \qquad (24)$$

which determines the evolution of the system. Now the problem of the time evolution of the cooperative spontaneous emission of three atoms becomes to find the eigenstates $|\Gamma_m\rangle$ and complex eigenvalues $\Gamma_m$ of $\mathbf{\Gamma}$. By solving a cubic equation, we can obtain the analytic expressions of three eigenvalues:

$$\begin{aligned}
\Gamma_a &= \Gamma_0 + \frac{\sqrt{3}}{3}\sqrt{\Gamma_{12}^2 + \Gamma_{13}^2 + \Gamma_{23}^2}\left(\cos\frac{\theta}{3} + \sqrt{3}\sin\frac{\theta}{3}\right) \\
\Gamma_b &= \Gamma_0 + \frac{\sqrt{3}}{3}\sqrt{\Gamma_{12}^2 + \Gamma_{13}^2 + \Gamma_{23}^2}\left(\cos\frac{\theta}{3} - \sqrt{3}\sin\frac{\theta}{3}\right), \\
\Gamma_c &= \Gamma_0 - \frac{2\sqrt{3}}{3}\sqrt{\Gamma_{12}^2 + \Gamma_{13}^2 + \Gamma_{23}^2}\cos\frac{\theta}{3}
\end{aligned} \qquad (25)$$

where $\theta = \arccos T$ with $T = -3\sqrt{3}\cdot \Gamma_{12}\Gamma_{13}\Gamma_{23}\cdot\left(\Gamma_{12}^2 + \Gamma_{13}^2 + \Gamma_{23}^2\right)^{-\frac{3}{2}}$. The expression of the eigenvectors is complicated, which can be obtained by numerical calculation. With the eigenvalues and eigenvectors, the time evolution of the state can be written as

$$|\psi(t)\rangle = \sum_m a_m e^{-\Gamma_m t}|\Gamma_m\rangle ,\qquad(26)$$

where $a_m$ is determined by the initial state,

$$|\psi(0)\rangle = \sum_{m=1}^{3} a_m |\Gamma_m\rangle .\qquad(27)$$

## . DECAY RATES AND LAMB SHIFT

After obtaining the eigenvalues of the matrix $\Gamma$, we can discuss the decay rates and the Lamb shift of the eigenstates. The three eigenvalues of the matrix can be written as:

$$\Gamma_{a,b,c} = \frac{\gamma_{a,b,c}}{2} + i\delta_{a,b,c} ,\qquad(28)$$

where $\gamma_{a,b,c}$ and $\delta_{a,b,c}$ are the decay rates and the collective Lamb shifts of the three eigenstates in long time limit. With Eq. (25) we can calculate the decay rates and Lamb shifts of the eigenstates for any distribution of atoms. Next, we consider two special cases.

### A. EQUILATERAL TRIANGLE

First, we consider the symmetry case: The three atoms are arranged as an equilateral triangle ($r_{mn} = r$) and the dipole moment is perpendicular to the plane of the three atoms ($\eta_{mn} = \eta = \pi/2$). In this case, the three atoms are equivalent and all the non-diagonal elements $\Gamma_{mn}$ are the same (denoted as $\Gamma_1$). The Eq. (25) can be simplified as:

$$\Gamma_a = \Gamma_0 + 2\Gamma_1 ,\quad \Gamma_b = \Gamma_c = \Gamma_0 - \Gamma_1 .\qquad(29)$$

Note that two eigenvalues are the same, so that the corresponding eigenstates $|b\rangle$ and $|c\rangle$ are not uniqueness. The three eigenstates can be written as

$$|a\rangle = \frac{1}{\sqrt{3}}(|e_1\rangle + |e_2\rangle + |e_3\rangle)$$
$$|b\rangle = \frac{1}{\sqrt{2}}(|e_2\rangle - |e_3\rangle) \qquad . \qquad (30)$$
$$|c\rangle = \frac{1}{\sqrt{6}}(|e_2\rangle + |e_3\rangle - 2|e_1\rangle)$$

Note that the eigenstate $|a\rangle$ is the Dicke state, $|D\rangle = \frac{1}{\sqrt{N}}\sum_n |e_n\rangle$ [7,10], of the three-atom system, where each atom has the same probability in its excited level with the same phase. From the matrix $\Gamma$, we can see that the Dicke state is an eigenstate of the system only when all the non-diagonal elements $\Gamma_{mn}$ are the same. This only happens in two cases: the system of two identical atoms and the system of three identical atoms in the equilateral triangle with the dipole moment perpendicular to the plane of the three atoms. For $N \geq 4$, the Dicke state can't exactly be the eigenstate, because there is no atomic space distribution resulting in the same interactions between any two atoms. For example, in the system of $N = 4$, the four atoms can be distributed as a regular tetrahedron, where the distance between any two atoms is the same, but the same angle $\eta_{mn}$ cannot be achieved, so that the non-diagonal elements will not equal, and the Dicke state will not be an eigenstate.

According to Eq. (28) and Eq. (29), we can obtain the decay rates:

$$\gamma_a = \gamma_{eg}\left[1 + 2D(x, \frac{\pi}{2})\right], \quad \gamma_b = \gamma_c = \gamma_{eg}\left[1 - D(x, \frac{\pi}{2})\right], \qquad (31)$$

where $x \equiv r/\lambda_0$. Note that the decay rates are modified by the function $D(x,\eta)$. From Fig. 2, we know that the interaction between atoms can greatly change the collective decay rates of the system, when the distance is approximately smaller than $\lambda_0/3$. With the atomic distance tends to zero, $D(x, \frac{\pi}{2})$ becomes one, and consequently, the decay rate of the Dicke state $|a\rangle$ will be increased and triple the decay rate of a single atom, while the decay rates of $|b\rangle$ and $|c\rangle$ tend to zero. Note that $D(x, \frac{\pi}{2})$ can be negative at certain distance (see Fig. 2(a)), and the Dicke state will not superradiant according to Eq. (31). At short atomic distance (less than about $\lambda_0/2$), it is superradiant. With the side-length expanding, it will oscillate between subradiant state and superradiant state.

According to Eq. (28) and Eq. (29), the Lamb shift of three eigenstates can be written as

$$\delta_a = \Delta_{eg} + \gamma_{eg} P(x, \frac{\pi}{2}), \qquad \delta_b = \delta_c = \Delta_{eg} - \frac{\gamma_{eg}}{2} P(x, \frac{\pi}{2}). \qquad (32)$$

Note that the Lamb shifts of $|b\rangle$ and $|c\rangle$ are the same, i.e., the two states are degenerated due to the symmetry. The energy difference between the symmetric state and asymmetric state is

$$\delta_a - \delta_b = \delta_a - \delta_c = \frac{3}{2} \gamma_{eg} P(x, \frac{\pi}{2}), \qquad (33)$$

which tells us that the splitting between the symmetric and asymmetric states is determined by $P(x, \frac{\pi}{2})$. The function $P(x, \frac{\pi}{2})$ describes the dipole-dipole interaction which arises from the real and virtual photon exchange between atoms. It has huge contribution to the Lamb shift when the distance between atoms is small. The Lamb shift will change their sign with the distance increasing, see Eq. (32) and Fig. 2(b).

**B. STRAIGHT LINE**

We consider the case that the three atoms are arranged in a straight line where the left two atoms are fixed and the third (right) atom stays at different position. According to Eqs. (25) and (28), we plot the decay rates of the three eigenstates $\gamma_{a,b,c}$ as a function of $x_{23}$ in Fig. 3. From the Figs. 3(a) and 3(b), we can see that the largest superradiance is not at $x_{23} = 0$ when $x_{12}$ is fixed. It means the intuitive idea that the superradiance should become strong with decreasing the distance between atoms is not always correct. Actually the superradiance also depends on the situation of atomic distribution.

For small atomic distance, the decay rate of the superradiant state reaches its maximum approximately at $x_{23} = x_{12}$, see Figs. 3(a) and 3(b). In other words, if the sample has good symmetry and $x_{12}$ is small, the strong superradiance will appear. The smaller the atomic distance is, the closer at $x_{23} = x_{12}$ the maximal decay rate happens. When the distance $x_{23} = x_{12}$ goes smaller and smaller, the maximal decay rate graduates to $3\gamma_{eg}$. For large $x_{12}$, the maximal decay rate is clearly not at $x_{23} = x_{12}$, see Figs. 3(c) and (d), which can be explained from the asymmetry of the matrix $\Gamma$ with the two functions $D(x,\eta)$ and $P(x,\eta)$. Actually, for large enough $x_{12}$, the maximal decay rate happens at $x_{23} \to 0$ (see the blue solid line in Fig. 3(d)) with the three-atom system reducing to the two-atom system.

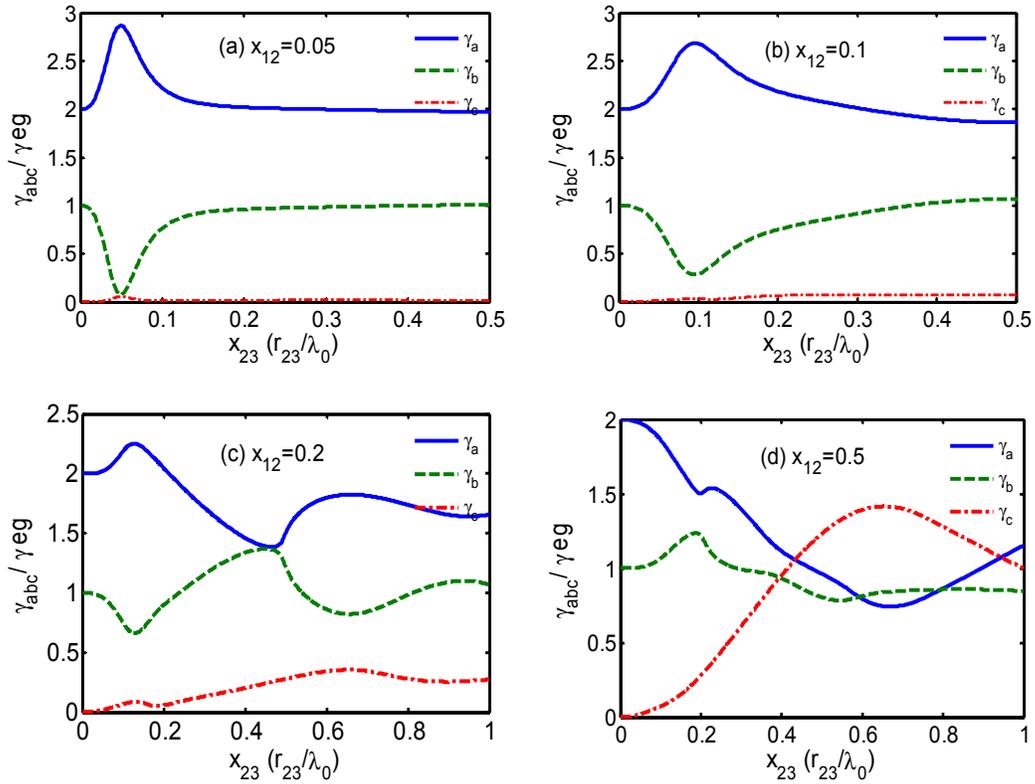

Fig.3 (color online) The decay rates of three eigenstates as a function of $x_{23}$ with different (a): $x_{12} = 0.05$; (b): $x_{12} = 0.1$; (c): $x_{12} = 0.2$; (d): $x_{12} = 0.5$. Here $\eta = \pi/2$.

## V. SPECTRUM

The probability of the detector absorbing a photon from the field at position $\mathbf{R}$ and time $t$ is proportional to the Glauber's first-order correlation function [18]

$$G_1(\mathbf{R},t) = \langle \psi_F | E^{(-)}(\mathbf{R},t) E^{(+)}(\mathbf{R},t) | \psi_F \rangle = \left| \langle 0 | E^{(+)}(\mathbf{R},t) | \psi_F \rangle \right|^2, \tag{34}$$

where $|\psi_F\rangle = \sum_{\mathbf{k}} C_{\mathbf{k}}(\infty) |1_{\mathbf{k}}\rangle$ is the final state. The value of $C_{\mathbf{k}}(\infty)$ can be obtained from Eq. (18).

$$C_{\mathbf{k}}(\infty) = -i \sum_n V_{\mathbf{k},eg} e^{-i\mathbf{k}\cdot\mathbf{r}_j} \int_0^\infty e^{-i(\omega'_{eg}-\omega_k)t'} C_n(t') dt', \tag{35}$$

where $C_n(t')$ is the time evolution of the amplitudes. For an initial state $|\psi(0)\rangle = \sum_m a_m |\Gamma_m\rangle$, the time evolution state is $|\psi(t)\rangle = \sum_m a_m e^{-\Gamma_m t} |\Gamma_m\rangle$, and consequently, we have

$$C_n(t) = \langle e_n | \psi(t) \rangle = \sum_m a_m e^{-\Gamma_m t} \langle e_n | \Gamma_m \rangle = \sum_m a_m b_n^{(m)} e^{-\Gamma_m t}. \tag{36}$$

Here $b_n^{(m)} = \langle e_n | \Gamma_m \rangle$ is the projection of the eigenstate $|\Gamma_m\rangle$ on the $n$th single-atom excited state $|e_n;0\rangle$. Substituting Eq. (36) into Eq. (35), we get

$$C_{\mathbf{k}}(\infty) = -i \sum_n V_{eg,\mathbf{k}} e^{-i\mathbf{k}\cdot\mathbf{r}_n} \int_0^\infty e^{-i(\omega_{eg}-\omega_k)t'} \sum_m a_m b_n^{(m)} e^{-\Gamma_m t'} dt' = -i \sum_n \sum_m V_{eg,\mathbf{k}} e^{-i\mathbf{k}\cdot\mathbf{r}_n} \frac{a_m b_n^{(m)}}{\Gamma_m - i(\omega_k - \omega_{eg})}. \tag{37}$$

Then we have

$$\begin{aligned}
\langle 0 | E^{(+)}(\mathbf{R},t) | \psi_F \rangle &= \langle 0 | \sum_{\mathbf{k}} \sqrt{\frac{\omega_k}{2\varepsilon_0 V}} b_{\mathbf{k}} \hat{e}_{\mathbf{k}} e^{-i\omega_k t + i\mathbf{k}\cdot\mathbf{R}} \sum_{\mathbf{k}} C_{\mathbf{k}}(\infty) |1_{\mathbf{k}}\rangle \\
&= \frac{-iV}{(2\pi)^3} \int d\Omega \int k^2 dk \sqrt{\frac{\omega_k}{2\varepsilon_0 V}} \cdot \hat{e}_{\mathbf{k}} e^{-i\omega_k t + i\mathbf{k}\cdot\mathbf{R}} \sum_n \sum_m V_{\mathbf{k},eg} e^{-i\mathbf{k}\cdot\mathbf{r}_n} \frac{a_m b_n^{(m)}}{\Gamma_m - i(\omega_k - \omega_{eg})} \\
&= \frac{-i}{2\varepsilon_0 (2\pi)^3} e^{-i\omega_k t} \omega_{eg} \int k^2 dk \int d\Omega \cdot \hat{e}_{\mathbf{k}} \cdot (\hat{e}_{\mathbf{k}} \cdot \hat{\mathbf{d}}_{eg}) e^{i\mathbf{k}\cdot\mathbf{R}_n} \frac{2\omega_{eg}}{\omega_{eg} + \omega_k} \sum_n \sum_m \frac{a_m b_n^{(m)}}{\Gamma_m - i(\omega_k - \omega_{eg})}
\end{aligned} \tag{38}$$

where $\mathbf{R}_n = \mathbf{R} - \mathbf{r}_n$.

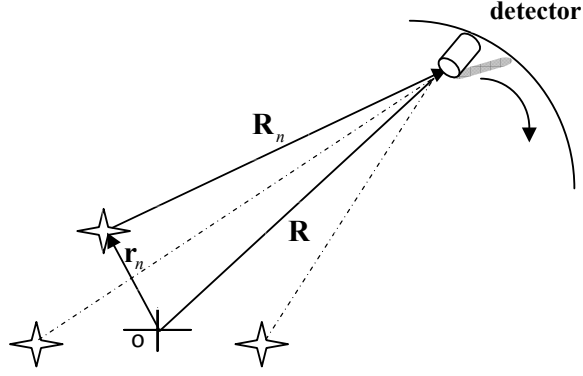

Fig. 4. The schematic description of the relation between the detector and the three atoms.

Note that the detector is far away from the atoms system in experiment ($R \gg \lambda_0$). The integral over the angle $\Omega$ of the EM modes can be written as

$$\int d\Omega \hat{e}_\mathbf{k} \cdot (\hat{e}_\mathbf{k} \cdot \hat{\mathbf{d}}_{eg}) e^{i\mathbf{k}\cdot\mathbf{R}_n} = \int d\Omega \left[ \hat{\mathbf{d}}_{eg} - \hat{\mathbf{k}}(\hat{\mathbf{k}} \cdot \hat{\mathbf{d}}_{eg}) \right] e^{i\mathbf{k}\cdot\mathbf{R}_n}$$
$$\approx \left[ \hat{\mathbf{d}}_{eg} - \hat{\mathbf{R}}_n (\hat{\mathbf{R}}_n \cdot \hat{\mathbf{d}}_{eg}) \right] \int_0^{2\pi} d\varphi \int_0^\pi e^{ikR_n \cos\theta} \sin\theta d\theta \qquad . \qquad (39)$$
$$\approx \left[ \hat{\mathbf{R}} \times (\hat{\mathbf{d}}_{eg} \times \hat{\mathbf{R}}) \right] \frac{2\pi}{ikR} (e^{ikR_n} - e^{-ikR_n})$$

The optical modes whose wavevectors is not parallel to $\hat{\mathbf{R}}_n$ lead to negligible contributions to the detector, so $\hat{\mathbf{k}}$ is replaced by $\hat{\mathbf{R}}_n$. Meanwhile, we use the approximation $R_n \approx R$ and $\hat{\mathbf{R}}_n \approx \hat{\mathbf{R}}$ except for the phases of the exponential function. Substituting Eq. (39) into Eq. (38), and only retaining the outward wave with the phase factor $e^{i(kR_n - \omega t)}$ [18], we get

$$\langle 0 | E^{(+)}(\mathbf{R},t) \sum_\mathbf{k} C_\mathbf{k}(\infty) | 1_\mathbf{k} \rangle$$
$$= \frac{-\omega_{eg} \left[ \hat{\mathbf{R}} \times (\hat{\mathbf{d}}_{eg} \times \hat{\mathbf{R}}) \right]}{i\varepsilon_0 (2\pi)^2 Rc^2} \int \omega_k d\omega_k \frac{\omega_{eg}}{\omega_{eg} + \omega_k} e^{-i\omega_k t} \sum_n \sum_m \frac{e^{ikR_n} a_m b_n^{(m)}}{\Gamma_m - i(\omega_k - \omega_{eg})} \quad , \qquad (40)$$
$$= \int d\omega_k e^{-i\omega_k t} B_\mathbf{R}(\omega_k)$$

$$B_\mathbf{R}(\omega_k) = \frac{\hat{\mathbf{R}} \times (\hat{\mathbf{d}}_{eg} \times \hat{\mathbf{R}})}{4i\varepsilon_0 \pi^2 Rc^2} \frac{\omega_{eg}^2 \omega_k}{\omega_{eg} + \omega_k} \sum_n \sum_m \frac{e^{ikR_n} a_m b_n^{(m)}}{\Gamma_m - i(\omega_k - \omega_{eg})} \quad . \qquad (41)$$

The spectrum detected by the detector at $\mathbf{R}$ is given by [5]

$$S_{\mathbf{R}}(\omega_k) \propto |B_{\mathbf{R}}(\omega_k)|^2 \propto \left| \hat{\mathbf{R}} \times (\hat{\mathbf{d}}_{eg} \times \hat{\mathbf{R}}) \sum_m \sum_n \frac{e^{ikR_n} a_m b_n^{(m)}}{\Gamma_m - i(\omega_k - \omega_{eg})} \right|^2, \quad (42)$$

which is the spectrum in a particular direction. Sometimes the total spectrum integrated over all detector direction is more important, which is the average of the spectra detected by all detectors in each direction [5].

Integrating the spectrum over all the detector direction, one obtains the total spectrum

$$S(\omega_k) = \int d\Omega_{\mathbf{R}} S_{\mathbf{R}}(\omega_k) = \sum_{m,n} f_n(k) f_m^*(k) T_{mn}(k), \quad (43)$$

where

$$f_n(k) = \sum_m \frac{a_m b_n^{(m)}}{\Gamma_m - i(\omega_k - \omega_{eg})}, \quad (44)$$

$$T_{mn}(k) = \int d\Omega_R \left[ \hat{\mathbf{R}} \times (\hat{\mathbf{d}}_{eg} \times \hat{\mathbf{R}}) \right]^2 e^{ikr_{mn}} = \begin{cases} 8\pi/3 & (m = n) \\ 4\pi D(x_{mn}, \theta_{mn}) & (m \neq n) \end{cases} \quad (45)$$

with $\theta_{mn}$ the angle between $\mathbf{r}_{mn}$ and $\hat{\mathbf{d}}_{eg}$ and $D(x_{mn}, \theta_{mn})$ the function defined in Eq. (22).

It is obvious that the spectrum depends on the initial state of the system. We assumed that we initially prepare the 1$^{th}$ atom in the excited state and the other two in the ground states: $|\psi_0\rangle = |e_1; 0\rangle$. By using Eq. (43), we can plot the spectrum of the three-atom system. First, we study the spectrum for the case (A) of the equilateral triangle. From Eqs. (32) and (42), we know that the spectrum will have two peaks at $\delta_k = \omega_k - \omega_{eg}^s = \gamma_{eg} P(x, \frac{\pi}{2})$ and $\delta_k = -\frac{\gamma_{eg}}{2} P(x, \frac{\pi}{2})$. In Fig. 5, we plot the total spectra of this special case (A). We can see that when the atoms are placed very near to each other (see the blue solid and green dashed lines in Fig. 5, where the side-length $x = 0.07, 0.1$), the interaction between the two atoms splits the spectrum into two peaks, the wide one (corresponding to superradiance) relates to the eigenstate $|a\rangle$ and the narrow one (subradiance) relates to the generated states $|b\rangle$ and $|c\rangle$. As the distance increases, the two peaks merge into one peak and finally, the spectrum tends to be the single-atom Lorentzian peak. If the three-atom system is initially prepared in the Dicke state, that is, the symmetric state $|a\rangle$ in this case, the spectrum will have only one peak.

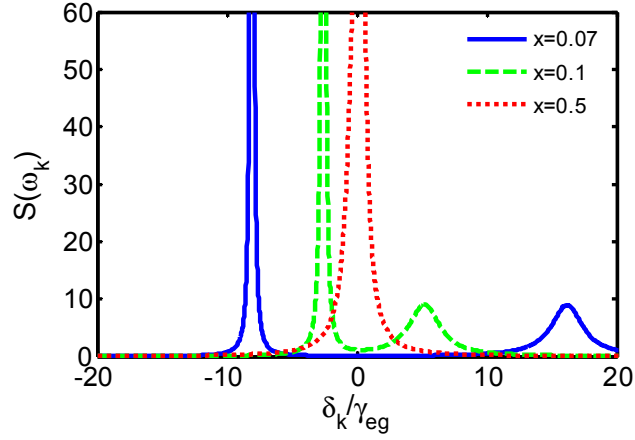

Fig. 5 (Color online). The total spectrum (in arbitrary units) of the initial state with only the first atom being excited in the special case that the three atoms are arranged as an equilateral triangle for different side-length ($x_{mn} = x$) and the dipole moment is perpendicular to the plane of the atoms ($\eta_{mn} = \eta = \pi/2$).

Next, we study the case (B) of a straight line. According to Fig. 2(b) and the results given by Ref. [6], we know that when the atomic distance is shorter than approximately $\lambda_0/3$, the two-atom system will have remarkable collective energy shift. So, we select the distance between the left two atoms at a small value ($x_{12} = 0.1$), then change the position of third one. In Fig. 6, we plot the total spectra for this case with (a) $\eta = \dfrac{\pi}{2}$ and (b) $\eta = 0$. When the three atoms are placed in short distance, the emission spectra have three peaks. The wide peak corresponds to the superradiant state, and the other two narrow ones come from two subradiant states. With the third (right) atom being put far away from the left two atoms along the line, the coherence between it and other two atoms weakens, so that the collective Lamb shift is receded. When the third atom is far enough away from the other two, the three-atom system reduces to the two-atom one with a two-peak spectrum

[5]. Moreover, in Fig. 6(a), when $\eta=\frac{\pi}{2}$, the superradiant state (corresponding to the widest peak) is blue-shifted while in Fig. 6(b), when $\eta=0$, the superradiant state is red-shifted. This $\eta$ dependence of the shift can be explained with the function $P(x,\eta)$ in Fig. 2(b). In short distance, $P(x,\eta=\frac{\pi}{2})$ stands for a repulsive interaction, while $P(x,\eta=0)$ stands for an attractive interaction. As a result, the superradiant state will have higher energy (blue-shift) when $\eta=\frac{\pi}{2}$, and lower energy (red-shift) when $\eta=0$.

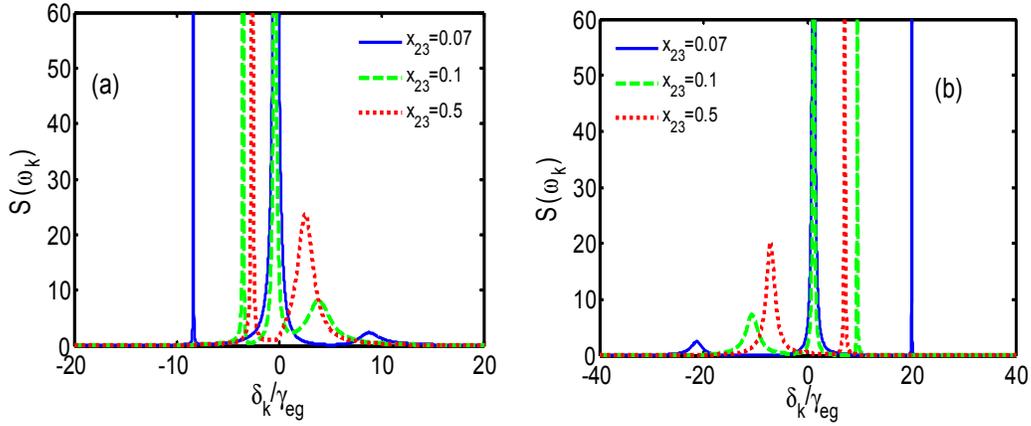

Fig. 6 (Color online). The total spectrum (in arbitrary units) for the case that three atoms are arranged in a straight line for different direction of dipole moment (a) $\eta_{mn}=\eta=\frac{\pi}{2}$ and (b) $\eta_{mn}=\eta=0$. Here $x_{12}=0.1$.

## VI. SUMMARY

In this paper, we studied the cooperative spontaneous emission of the system of three identical atoms in vacuum. Focusing on the single-atom-excitation subspace, we investigated the eigenmodes as well as the corresponding collective decay rates and Lamb shifts for the three-atom system under different atomic distributions.

For the case that the three atoms are arranged as an equilateral triangle with the dipole moment perpendicular to the plane of the triangle, the symmetric Dicke state is shown to be an "exponentially-decaying" eigenstate of the system. In this case, spontaneous emission of the Dicke state can be superradiant or subradiant, depending on the distance between atoms.

For the case of the three atoms in a line, the Dicke state is no longer the engenstate of the system and the superradiance of the system is not always enhanced with the distance $x_{23}$ decreasing (fixing $x_{12}$). We find that the superradiant limit $3\gamma_{eg}$ will be obtained when $x_{23} \approx x_{12}$ and they are short enough (approximately less than $0.1\lambda_0$). It means the symmetry of atoms' distribution have larger influence on the superradiance of the system.

We also studied the spontaneous emission spectra of the initial state with only the first atom being excited. Such an initial state is composed of the three eigenstates and usually the spectrum would have three peaks. In the first case of equilateral triangle, the spectrum has two peaks due to the two-fold degeneracy of the asymmetric states, resulting from the symmetry distribution of the three atoms. In the second case of straight line, the spectrum has three peaks due to no degeneracy, and the collective Lamb shift is dependent on the angle $\eta$ (between the dipole moment and the straight line). When $\eta = \pi/2$, the collective Lamb shift of superradiant state is blue-shifted, while, when $\eta = 0$, the collective Lamb shift of superradiant state is red-shifted.

**Acknowledgments**

We thank Jörg Evers for helpful discussions. This work was supported by the National Natural Science Foundation of China (under Grant Nos. 11174026, 11174027) and the National Basic Research Program of China (No. 2011CB922203 and No. 2012CB922104).

**Appendix**: Calculation of the Matrix Elements in Eq. (19)

Formally integrating Eq. (18) with extending the lower bound of time integration to $-\infty$ and replacing $C_n(t')$ with $C_n(t)$, we get

$$C_{\mathbf{k}}(t) = -i \sum_n V_{\mathbf{k},eg} e^{-i\mathbf{k}\cdot\mathbf{r_n}} C_n(t) \int_{-\infty}^{t} e^{-i(\omega'_{eg}-\omega_k)t'} dt'. \tag{A1}$$

Substituting Eq. (A1) into Eq. (17), one has

$$\begin{aligned}
\dot{C}_n(t) &\approx -\sum_{\mathbf{k}} V_{\mathbf{k},eg}^2 \int_{-\infty}^{t} e^{i(\omega'_{eg}-\omega_k)(t-t')} dt' \cdot C_n(t) \\
&\quad - \sum_{m \neq n} \left[ \sum_{\mathbf{k}} V_{\mathbf{k},eg}^2 e^{i k r_{nm}} \int_{-\infty}^{t} e^{i(\omega'_{eg}-\omega_k)(t-t')} dt' + i\Delta_{sta}^{mn} \right] \cdot C_m(t), \\
&= -\Gamma_0 \cdot C_n(t) - \sum_{m \neq n} \Gamma_{mn} \cdot C_m(t)
\end{aligned} \tag{A2}$$

where

$$\Gamma_0 = \sum_{\mathbf{k}} V_{\mathbf{k},eg}^2 \left[ \pi\delta(\omega'_{eg}-\omega_k) + i\wp \frac{1}{\omega'_{eg}-\omega_k} \right], \tag{A3}$$

$$\Gamma_{mn} = \sum_{\mathbf{k}} V_{\mathbf{k},eg}^2 e^{i\mathbf{k}\cdot\mathbf{r}_{nm}} \left[ \pi\delta(\omega'_{eg}-\omega_k) \right] + i\left( \Delta_{sta}^{mn} + \wp \sum_{\mathbf{k}} V_{\mathbf{k},eg}^2 e^{i\mathbf{k}\cdot\mathbf{r}_{nm}} \frac{1}{\omega'_{eg}-\omega_k} \right). \tag{A4}$$

In the above equation, we have used the integral formula $\int_{-\infty}^{t} dt' e^{ix(t-t')} = \pi\delta(x) + i\wp \frac{1}{x}$, and $\wp$ stands for principle value.

Detailed calculation gives

$$\begin{aligned}
\Gamma_0 &= \sum_{\mathbf{k}} V_{\mathbf{k},eg}^2 \left( \pi\delta(\omega'_{eg}-\omega_k) + i\wp \frac{1}{\omega'_{eg}-\omega_k} \right) \\
&= \frac{d_{eg}^2 \omega_{eg}^2}{6\pi^2 \varepsilon_0 c^3} \int_0^{\infty} \omega_k \left( \pi\delta(\omega'_{eg}-\omega_k) + i\wp \frac{1}{\omega'_{eg}-\omega_k} \right) d\omega_k \\
&\approx \frac{d_{eg}^2 \omega_{eg}^2}{6\pi^2 \varepsilon_0 c^3} \int_0^{\infty} \omega_k \left( \pi\delta(\omega_{eg}-\omega_k) + i\wp \frac{1}{\omega_{eg}-\omega_k} \right) d\omega_k, \\
&= \frac{1}{2} \cdot \frac{d_{eg}^2 \omega_{eg}^3}{3\pi \varepsilon_0 c^3} + \frac{d_{eg}^2 \omega_{eg}^2}{6\pi^2 \varepsilon_0 c^3} \wp \int_0^{\infty} \frac{\omega_k}{\omega_{eg}-\omega_k} \frac{4\omega_{eg}^2}{(\omega_k+\omega_{eg})^2} d\omega_k \\
&= \frac{\gamma_{eg}}{2} + i\Delta_{eg}
\end{aligned} \tag{A5}$$

where

$$\Delta_{eg} = \frac{d_{eg}^2 \omega_{eg}^2}{6\pi^2 \varepsilon_0 c^3} \wp \int_0^{\infty} \frac{\omega_k}{\omega_{eg}-\omega_k} \frac{4\omega_{eg}^2}{(\omega_k+\omega_{eg})^2} d\omega_k \tag{A6}$$

is the dynamic shift of a single atom.

The real part of $\Gamma_{mn}$ is

$$\mathrm{Re}(\Gamma_{mn}) = \sum_{\mathbf{k}} V_{\mathbf{k},eg}^2 e^{i\mathbf{k}\cdot\mathbf{r}_{mn}} \left(\pi\delta(\omega_{eg} - \omega_k)\right)$$

$$= \frac{d_{eg}^2 \omega_{eg}^2}{6\pi\varepsilon_0 c^3} \int_0^\infty \left[\frac{2\omega_{eg}}{(\omega_{eg} + \omega_k)}\right]^2 \omega_k D(kr_{mn}/2\pi, \eta_{mn}) \delta(\omega_{eg} - \omega_k) d\omega_k \qquad (A7)$$

$$= \frac{d_{eg}^2 \omega_{eg}^3}{6\pi\varepsilon_0 c^3} D(x_{mn}, \eta_{mn})$$

$$= \frac{\gamma_{eg}}{2} D(x_{mn}, \eta_{mn})$$

The imaginary part of $\Gamma_{mn}$ is

$$\mathrm{Im}(\Gamma_{mn}) = \left(\Delta_{sta}^{mn} + \wp \sum_{\mathbf{k}} V_{\mathbf{k},eg}^2 e^{i\mathbf{k}\cdot\mathbf{r}_{mn}} \frac{1}{\omega'_{eg} - \omega_k}\right)$$

$$= \frac{1}{4\pi\varepsilon_0}\left[\frac{\mathbf{d}_{eg}^{(m)}\cdot\mathbf{d}_{ge}^{(n)}}{r_{mn}^3} - \frac{3(\mathbf{d}_{eg}^{(m)}\cdot\mathbf{r}_{mn})(\mathbf{d}_{ge}^{(n)}\cdot\mathbf{r}_{mn})}{r_{mn}^5}\right] - \sum_{\mathbf{k}} \frac{2g_{\mathbf{k},eg}^2 \xi_{k,eg}}{\omega_k}(2-\xi_{k,eg})e^{i\mathbf{k}\cdot\mathbf{r}_{mn}} + \wp\sum_{\mathbf{k}} V_{\mathbf{k},eg}^2 e^{i\mathbf{k}\cdot\mathbf{r}_{mn}} \frac{1}{\omega'_{eg} - \omega_k}$$

$$= \frac{1}{4\pi\varepsilon_0}\left[\frac{\mathbf{d}_{eg}^{(m)}\cdot\mathbf{d}_{ge}^{(n)}}{r_{mn}^3} - \frac{3(\mathbf{d}_{eg}^{(m)}\cdot\mathbf{r}_{mn})(\mathbf{d}_{ge}^{(n)}\cdot\mathbf{r}_{mn})}{r_{mn}^5}\right] - \frac{d_{eg}^2 \omega_{eg}^2}{6\pi^2\varepsilon_0 c^3}\int_0^\infty \frac{2(\omega_k + 2\omega_{eg})\omega_k}{(\omega_{eg}+\omega_k)^2} D(kr_{mn}/2\pi, \eta_{mn})d\omega_k$$

$$+ \frac{d_{eg}^2 \omega_{eg}^2}{6\pi^2\varepsilon_0 c^3}\wp\int_0^\infty \frac{\omega_k}{(\omega_{eg}-\omega_k)}\frac{4\omega_{eg}^2}{(\omega_{eg}+\omega_k)^2} D(kr_{mn}/2\pi, \eta_{mn})d\omega_k$$

$$= \frac{1}{4\pi\varepsilon_0}\left[\frac{\mathbf{d}_{eg}^{(m)}\cdot\mathbf{d}_{ge}^{(n)}}{r_{mn}^3} - \frac{3(\mathbf{d}_{eg}^{(m)}\cdot\mathbf{r}_{mn})(\mathbf{d}_{ge}^{(n)}\cdot\mathbf{r}_{mn})}{r_{mn}^5}\right] + \frac{d_{eg}^2 \omega_{eg}^2}{6\pi^2\varepsilon_0 c^3}\wp\int_0^\infty \frac{2\omega_k^2}{\omega_{eg}^2 - \omega_k^2} D(kr_{mn}/2\pi, \eta_{mn})d\omega_k$$

$$\approx \frac{d_{eg}^2}{4\pi\varepsilon_0 r_{mn}^3}(1-3\cos^2\eta_{mn}) + \frac{\gamma_{eg}}{2}\frac{3}{2}\left\{-\sin^2\eta_{mn}\cdot\frac{\cos(2\pi x_{mn})}{2\pi x} + (1-3\cos^2\eta)\left[\frac{\sin(2\pi x_{mn})}{(2\pi x_{mn})^2} + \frac{\cos(2\pi x_{mn})-1}{(2\pi x_{mn})^3}\right]\right\}$$

$$= \frac{\gamma_{eg}}{2}\frac{3}{2}\left\{-\sin^2\eta_{mn}\cdot\frac{\cos(2\pi x_{mn})}{2\pi x_{mn}} + (1-3\cos^2\eta_{mn})\left[\frac{\sin(2\pi x_{mn})}{(2\pi x_{mn})^2} + \frac{\cos(2\pi x_{mn})}{(2\pi x_{mn})^3}\right]\right\}$$

$$= \frac{\gamma_{eg}}{2} P(x_{mn}, \eta_{mn}).$$

In the above equations, $D(x_{mn}, \eta_{mn})$ and $P(x_{mn}, \eta_{mn})$ are defined in Eqs. (22) and (23) respectively. The method for changing the summation over $\mathbf{k}$ to the integration can be found in Ref. [4] (pp.7).